\documentclass[12pt,a4paper]{iopart}

\usepackage{graphicx}
\usepackage{color}
\usepackage{mathrsfs} 
\usepackage{dsfont}
\usepackage{xspace}
\usepackage{iopams}  

\usepackage[bookmarks=true,
            bookmarksopen=true,colorlinks=false]{hyperref}

\newcommand{\eqref}{\eref}
\newcommand{\dd}{\mathrm{d}}
\newcommand{\ee}{\mathrm{e}}
\newcommand{\ii}{\mathrm{i}}

\newcommand{\CC}{\mathds C}

\newcommand{\E}{\mathcal E}

\newcommand{\A}{\mathcal A}
\renewcommand{\H}{{\mathcal H}}

\newcommand{\Y}{{\mathbf Y}}
\newcommand{\J}{{\mathbf J}}
\newcommand{\B}{{\mathbf B}}
\newcommand{\M}{{\mathbf M}}
\newcommand{\N}{{\mathbf N}}
\renewcommand{\P}{{\mathbf P}}
\newcommand{\Cmat}{{\mathbf C}}
\newcommand{\Dmat}{{\mathbf D}}
\newcommand{\1}{{\mathbf 1}}
\newcommand{\0}{{\mathbf 0}}
\newcommand{\EE}{{\mathbf E}}
\renewcommand{\Re}{\mathrm{Re}}
\renewcommand{\Im}{\mathrm{Im}}

\begin{document}

\title
{Axis potentials for stationary $n$-black-hole configurations}

\author{J\"org Hennig}
\address{Department of Mathematics and Statistics,
           University of Otago,
           PO Box 56, Dunedin 9054, New Zealand}
\eads{\mailto{jhennig@maths.otago.ac.nz}}

\begin{abstract}
We extend earlier discussions of the balance problem for two black holes and study stationary spacetimes containing an arbitrary number of $n$ aligned rotating and (possibly) charged black holes. For these hypothetical equilibrium configurations, we obtain the most general form of the boundary data on the symmetry axis in terms of a finite number of parameters. Hence future investigations of $n$-black-hole configurations can be restricted to studying properties of these finite families of solutions.
 \\[2ex]{}
\end{abstract}

\section{Introduction\label{sec:intro}}

An intriguing open problem in general relativity is the question as to whether stationary equilibrium configurations with clearly separated bodies can exist. Due to the nonlinear effect of the \emph{spin--spin repulsion} of rotating objects, and perhaps by considering charged objects with an additional \emph{electromagnetic repulsion}, it remains a possibility that such unusual configurations do exist. Any definite and general answer to this question --- whether in the affirmative or negative --- would provide important insights into the nature of the gravitational interaction in general relativity.

While a solution to the balance problem in its most general form --- with an arbitrary number of objects of an arbitrary type --- is probably far out of reach, partial results were obtained in the special case of two-black-hole configurations. If we consider two \emph{uncharged} black holes in vacuum, then it is known that there is no physically acceptable equilibrium configuration \cite{NeugebauerHennig2009,HennigNeugebauer2011,NeugebauerHennig2012,Chrusciel2011}. The corresponding proof roughly proceeds as follows. Firstly, it was shown that, if any regular solutions existed, they would necessarily be members of the well-known \emph{double-Kerr-NUT} family of solutions, first derived in \cite{KramerNeugebauer1980,Neugebauer1980}. Secondly, it turned out that for all possible candidate solutions, at least one of the black holes would violate an inequality between its angular momentum and horizon area \cite{HennigAnsorgCederbaum2008} that holds for all regular black holes. Consequently, there cannot be any such regular configurations in vacuum.

A first step towards a generalisation of this result to \emph{charged} two-black-hole configurations in electrovacuum was made in \cite{Hennig2019}. By studying a boundary value problem for the Einstein--Maxwell equations, it was shown that any possible equilibrium configuration is necessarily in a family of solutions that depends on a small number of parameters and is characterised by rational axis data (for the Ernst potentials) of a certain form. However, it remains to be determined whether this family contains any physically acceptable solution that is free from naked singularities and has an appropriate behaviour both on the symmetry axis and at infinity.\footnote{A well-defined \emph{mathematical} solution to the balance problem with charged black holes is, of course, given by the \emph{Majumdar-Papapetrou solution} \cite{Majumdar1947,Papapetrou1947}, which describes the superposition of even an arbitrary number of extremal Reissner-Nordstr\"om black holes. However, since it is thought that extremal black holes with degenerate horizons can only be approached, but not exactly reached, in nature, physically relevant solutions should be \emph{non-extremal}.}

Here, we demonstrate that the considerations from \cite{Hennig2019} can be extended to configurations with an \emph{arbitrary} number of $n$ charged and rotating black holes, and we construct the most general form of the axis potentials. Therefore, for each $n$, we uniquely characterise a finite solution family that contains all solutions to the corresponding balance problem --- if any exist.

\section{The linear problem\label{sec:LP}}

\begin{figure}\centering
 \includegraphics[width=6.3cm]{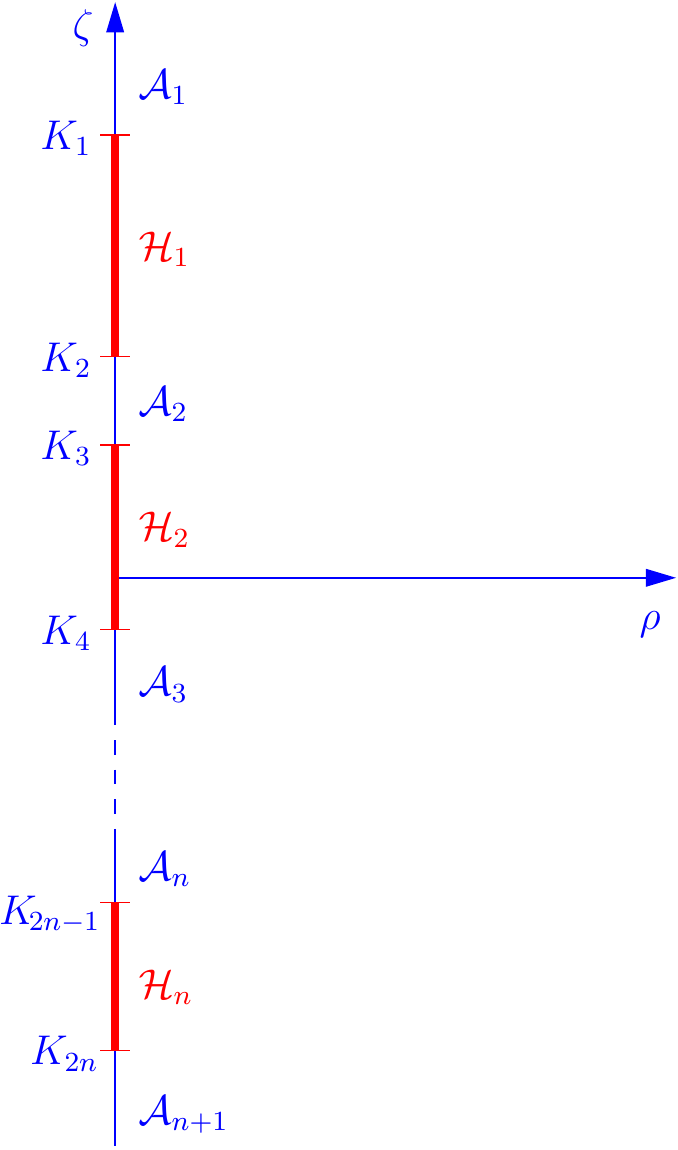}
 \caption{Illustration of an $n$-black-hole equilibrium configuration in Weyl-Lewis-Papapetrou coordinates.\label{fig:nBHs}}
\end{figure}

We consider (hypothetical) $n$-black-hole configurations that consist of $n$ aligned black holes. The black holes are (potentially) charged, and they rotate with angular velocities $\Omega_i\neq0$, $i=1,\dots,n$. In Weyl-Lewis-Papapetrou coordinates $(\rho,\zeta,\varphi,t)$,
the event horizons $\H_1$, $\dots$, $\H_n$ correspond to intervals on the $\zeta$-axis, and we denote the endpoints of these intervals by $K_1$, $K_2$; $K_3$, $K_4$; $\dots$; $K_{2n-1}$, $K_{2n}$, see Fig.~\ref{fig:nBHs}. The remaining parts $\A_1$, $\dots$, $\A_{n+1}$ of the $\zeta$-axis shown in the figure correspond to the symmetry axis.

The electrovacuum exterior of the black holes can be described with the line element
\begin{equation}\label{eq:metric}
 \dd s^2=f^{-1}
          \left[\ee^{2k}(\dd\rho^2+\dd\zeta^2)+\rho^2\,\dd\varphi^2\right]
         -f(\dd t+a\,\dd\varphi)^2,
\end{equation}
where $f$, $k$ and $a$ are functions of $\rho$ and $\zeta$ alone. The corresponding Einstein--Maxwell equations can be given in form of the well-known \emph{Ernst equations} for the complex Ernst potentials $\E(\rho,\zeta)$ and $\Phi(\rho,\zeta)$ \cite{Ernst1968b}.
Our entire solution procedure then relies on the remarkable fact that these, in turn, are equivalent to a linear matrix problem via its integrability condition \cite{Belinski1979, NeugebauerKramer}. We will use the version of Meinel \cite{Meinel2012}, where the linear problem (LP) consists in the following equations for a $3\times 3$ matrix function $\Y=\Y(\rho,\zeta;K)$ depending on the coordinates and the \emph{spectral parameter} $K\in\CC$,
\begin{eqnarray}
\label{eq:LP1}
 \Y_{,z} &=& \left[
 \left(\begin{array}{ccc}
   B_1 & 0 & C_1\\ 0 & A_1 & 0\\ D_1 & 0 & 0
 \end{array}\right)
 +\lambda
 \left(\begin{array}{ccc}
  0 & B_1 & 0\\ A_1 & 0 & -C_1\\ 0 & D_1 & 0
 \end{array}\right)\right]\Y,\\
 \label{eq:LP2}
 \Y_{,\bar z} &=& \left[
 \left(\begin{array}{ccc}
   B_2 & 0 & C_2\\ 0 & A_2 & 0\\ D_2 & 0 & 0
 \end{array}\right)
 +\frac{1}{\lambda}
 \left(\begin{array}{ccc}
  0 & B_2 & 0\\ A_2 & 0 & -C_2\\ 0 & D_2 & 0
 \end{array}\right)\right]\Y.
\end{eqnarray}
Here, we have introduced the complex coordinates $z=\rho+\ii\zeta$, $\bar z=\rho-\ii\zeta$ and the function 
\begin{equation}\label{eq:lambda}
 \lambda=\sqrt{\frac{K-\ii\bar z}{K+\ii z}}.
\end{equation}
The matrix elements in \eqref{eq:LP1}, \eqref{eq:LP2} are given by
\begin{eqnarray}
 A_1 &= \bar B_2 = \frac{1}{2f}(\E_{,z}+2\bar\Phi\Phi_{,z}),\quad
 C_1 &= f\bar D_2 = \Phi_{,z},\\
 A_2 &= \bar B_1 = \frac{1}{2f}(\E_{,\bar z}+2\bar\Phi\Phi_{,\bar z}),\quad
 C_2 &= f\bar D_1 = \Phi_{,\bar z},
\end{eqnarray}
where the metric function $f$ can be expressed in terms of the Ernst potentials as $f=\Re(\E)+|\Phi|^2$.

Note that the square-root function $\lambda$ and, consequently, the solution to the LP, are defined on a two-sheeted Riemannian $K$-surface. As discussed in \cite{Hennig2019}, the solutions $\Y$ on the two sheets are connected via the simple relation
\begin{equation}\label{eq:sheets}
  \Y|_{-\lambda}=\J\Y|_{\lambda}\B,\quad
  \J:=\mathrm{diag}(1,-1,1)
\end{equation}
with some matrix function $\B=\B(K)$, which has the property that $\B^2$ gives the $3\times3$ identity matrix, $\B^2=\1$.

Finally, it is most useful to also consider the LP in reference frames that are co-rotating with one of the black holes. These are obtained with the transformation $\varphi\mapsto\tilde\varphi=\varphi-\Omega t$, where $\Omega$ is chosen to coincide with one of the parameters $\Omega_i$.
The solution $\tilde\Y$ to the LP in the new frame can easily be obtained from $\Y$ via
\cite{AnsorgHennig2009,HennigAnsorg2009,Meinel2012}
\begin{equation}\fl\label{eq:rotframe}
 \tilde\Y(\rho,\zeta; K)=\left[
  \left(\begin{array}{ccc}
         c_- & 0   & 0\\
         0   & c_+ & 0\\
         0   & 0   & 1
        \end{array}\right)
  +\ii(K+\ii z)\frac{\Omega}{f}
  \left(\begin{array}{ccc}
         -1      & -\lambda & 0\\
         \lambda & 1        & 0\\
         0       & 0        & 0
        \end{array}\right)
\right]\Y(\rho,\zeta;K),
\end{equation}
where
\begin{equation}
 c_{\pm}:=1+\Omega\left(a\pm\frac{\rho}{f}\right).
\end{equation}

Now we have all necessary ingredients to study the LP at the symmetry axis, the horizons and at infinity, which will allow us to explicitly construct the most general axis potentials on the upper axis part $\A_1$.

\section{Construction of the axis potentials\label{sec:axispots}}

At $\rho=0$, i.e.\ at the symmetry axis and horizons, the LP reduces to an ODE that can be solved exactly. Note that the two Riemannian $K$-sheets are characterised by $\lambda=\pm1$ at $\rho=0$, cf.~\eqref{eq:lambda}. In the sheet with $\lambda=1$, we obtain (using the usual axis boundary conditions \cite{Hennig2019}) for the solutions $\Y$ in the non-rotating and $\tilde\Y$ in the rotating reference frames
\begin{eqnarray}
 \fl\label{eq:axisY}
 \A_j\ (j=1,\dots,n+1):\ && \Y = \EE\Cmat_j,\quad
       \EE:=\left(\begin{array}{ccc}
             \bar\E+2|\Phi|^2 & 1 & \Phi\\
             \E & -1 & -\Phi\\
             2\bar\Phi & 0 & 1
            \end{array}\right),\\
  \fl
       &&\tilde\Y=\left[\EE+2\ii\Omega(K-\zeta)\left(\begin{array}{ccc}
                                                      -1 & 0 & 0\\
                                                      1 & 0 & 0\\
                                                      0 & 0 & 0
                                                     \end{array}\right)\right]\Cmat_j,\\
  \fl
 \H_j(j=1,\dots,n):\ && \Y=\EE\Dmat_j,\\
           && \tilde\Y=\left[\left(\begin{array}{ccc}
                               0 & 0 & 0\\ 0 & 0 & 0\\ 0 & 0 & 1
                              \end{array}\right)
\EE+2\ii\Omega(K-\zeta)\left(\begin{array}{ccc}
                                                      -1 & 0 & 0\\
                                                      1 & 0 & 0\\
                                                      0 & 0 & 0
                                                     \end{array}\right)\right]\Dmat_j,
\end{eqnarray}
where the $3\times3$ matrices $\Cmat_j=\Cmat_j(K)$ and $\Dmat_j=\Dmat_j(K)$ are integration `constants'.
In the following, we will choose the rotating frames with $\Omega=\Omega_j$ on $\H_j$, and $\Omega=\Omega_{j-1}$ or $\Omega=\Omega_j$ on $\A_j$ (corresponding to the angular velocities of the two horizons adjacent to $\A_j$). For a compact notation of some of the following equations, we also introduce the quantities
\begin{equation}
 \omega_1=\omega_2=\Omega_1,\
 \omega_3=\omega_4=\Omega_2,\ \dots,\
 \omega_{2n-1}=\omega_{2n}=\Omega_n,
\end{equation}
such that a horizon with an endpoint at $\zeta=K_j$ has the angular velocity $\omega_j$.

In a next step, we observe that the above matrices $\Cmat_j$ and $\Dmat_j$ cannot be chosen independently. Instead, they are restricted by the requirement of continuity of $\Y$ and $\tilde\Y$ at $\zeta=K_j$, where the symmetry axis parts and horizons meet. The resulting algebraic conditions lead to the following relations,
\begin{equation}\fl \label{eq:CDrelations}
 \Dmat_j = \left(\1+\frac{1}{\alpha_{2j-1}}\M_{2j-1}\right)\Cmat_j,\quad
 \Cmat_{j+1} = \left(\1-\frac{1}{\alpha_{2j}}\M_{2j}\right)\Dmat_j,\quad
 j=1,\dots,n,
\end{equation}
where, for $k=1,\dots,2n$, we have defined the constants
\begin{equation}\label{eq:alphadef}
 \alpha_k:=2\ii\omega_k(K-K_k),
\end{equation}
and the nilpotent matrices ($\M_k^2=\0$)
\begin{equation}\fl
 \M_k:=\left(\begin{array}{c}
       -1\\ \bar\E_k\\ 2\bar\Phi_k
      \end{array}\right)
   \left(\begin{array}{ccc}
             -\E_k & 1 & \Phi_k
            \end{array}\right),\quad
 \E_k:=\E(0,K_k),\quad \Phi_k:=\Phi(0,K_k).
\end{equation}

Furthermore, we also have to ensure the correct transition from the upmost axis part $\A_1$ to the lowest axis part $A_{n+1}$ along an infinitely large semicircle. As discussed in \cite{Hennig2019}, the required fall-off behaviour of the Ernst potentials in an asymptotically flat spacetime implies that the solution to the LP `at infinity' is independent of the coordinates $\rho$ and $\zeta$ and can only depend on $K$. Moreover, if the semicircle is parametrised by an angle $0\le\alpha\le\pi$, then we have $\lambda=\pm\ee^{\ii\alpha}$ there. Hence we have to connect the $\lambda=1$ solution on $\A_1$ with the $\lambda=-1$ solution on $\A_{n+1}$ (and vice versa). Together with \eqref{eq:sheets}, this leads to the additional condition
\begin{equation}\label{eq:infrelation}
 \Cmat_{n+1}\B = \P\Cmat_1,\quad
 \P:=\left(\begin{array}{ccc}
            0 & 1 & 0\\
            1 & 0 & 0\\
            0 & 0 & 1
           \end{array}\right).
\end{equation}

The previous relations can be used to derive \emph{parameter conditions} --- algebraic equations that restrict the allowed values for the quantities  $\E_1$, $\dots$, $\E_{2n}$, $\Phi_1$, $\dots$, $\Phi_{2n}$, $K_1$, $\dots$, $K_{2n}$, and $\Omega_1$, $\dots$, $\Omega_n$. For that purpose, we combine all equations in \eqref{eq:CDrelations} with \eqref{eq:infrelation} such that all matrices $\Cmat_j$ and $\Dmat_j$ with exception of $\Cmat_1$ are eliminated. The resulting equation can be solved for $\B$,
\begin{equation}\label{eq:Bformula}
 \B = \Cmat_1^{-1}\left[\prod_{j=1}^{2n} \Big(\1+\frac{(-1)^j}{\alpha_j}\M_j\Big)\right]\P\Cmat_1.
\end{equation}
Using $\B^2=\1$, the square of this formula is independent of $\B$, and $\Cmat_1$ cancels as well. Rearranging the result, we finally obtain
\begin{equation}
 \P\left[\prod_{j=2n}^1 \Big(\alpha_j\1-(-1)^j\M_j\Big)\right]
 -\left[\prod_{j=1}^{2n} \Big(\alpha_j\1+(-1)^j\M_j\Big)\right]\P=\0,
\end{equation}
where the matrix products are to be calculated from left to right in the indicated order (i.e.\ starting with the $j=2n$ term in the first product and with the $j=1$ term in the second product). With the definition \eqref{eq:alphadef} of $\alpha_j$, we observe that the left-hand side is a matrix polynomial in $K$ of degree $2n-1$ (because the terms proportional to $K^{2n}$ in the first and second product cancel). The requirement that all coefficients in this polynomial vanish leads to the desired parameter conditions.

For example, vanishing of the $K^{2n-1}$-term gives
\begin{equation}
 \sum_{j=1}^{2n}\left[\frac{(-1)^j}{\omega_j}(\P\M_j+\M_j\P)\right]=\0.
\end{equation}
In particular, if we consider the $1$-$1$ component, using $(\P\M_j+\M_j\P)_{11}=-1-|\E_j|^2$, we obtain the condition
\begin{equation}\label{eq:parcond}
 \frac{|\E_1|^2-|\E_2|^2}{\Omega_1} + \frac{|\E_3|^2-|\E_4|^2}{\Omega_2} +\dots
 +\frac{|\E_{2n-1}|^2-|\E_{2n}|^2}{\Omega_n}=0.
\end{equation}
Similarly, many more conditions can be derived from the other components and the coefficients of other powers of $K$.

Now we can generalise the procedure from \cite{Hennig2019} to the case of an arbitrary number of $n$ black holes and construct the axis potentials $\E(0,\zeta)$, $\Phi(0,\zeta)$ on the upmost axis part $\A_1$. As discussed above, the function $\lambda$ introduces a 2-sheeted Riemannian surface on which the solution $\Y$ to the LP is defined. The map $K\mapsto\lambda$ does have two branch points at $K_1=\zeta+\ii\rho$ and $K_2=\zeta-\ii\rho$, which become confluent branch points at $K=\zeta$ at the $\zeta$-axis ($\rho=0$). Since the two Riemannian sheets are connected at the branch points, the values of $\Y$ in the two sheets must coincide there.

On the axis part $\A_1$, we obtain the condition that $\Y$ takes on the same value at $K=\zeta$ in both sheets $\lambda=+1$ and $\lambda=-1$. With \eqref{eq:axisY} and \eqref{eq:sheets} this leads to
\begin{equation}\label{eq:sheetcond}
 \EE\Cmat_1=\J\EE\Cmat_1\B\quad\mbox{at}\quad K=\zeta.
\end{equation}
Similarly to \cite{Hennig2019}, we introduce  the matrix 
$\N(\zeta):=\EE^{-1}\J\EE\P|_{K=\zeta}$. With the definition of $\EE$ in \eqref{eq:axisY}, this can, on the one hand, be expressed in terms of the axis potentials on $\A_1$,
\begin{equation}\label{eq:defN}
 \N=
 \frac{1}{f}\left(\begin{array}{ccc}
                    1                  & |\Phi|^2-\ii b  & \Phi\\
                    |\Phi|^2+\ii b & |\E|^2            & -\Phi\bar\E\\
                    -2\bar\Phi       & 2\bar\Phi\E     & f-2|\Phi|^2
                   \end{array}\right)\quad\mbox{at}\quad \rho=0,\ \zeta\ge K_1,
\end{equation}
where $b=\Im(\E)$. On the other hand, using \eqref{eq:sheetcond} and \eqref{eq:Bformula}, we also obtain this representation for $\N$,
\begin{equation}\fl\label{eq:Nformula}
 \N = \prod_{j=1}^{2n} \Big(\1+\frac{(-1)^j}{\alpha_j}\M_j\Big)\Big|_{K=\zeta}
 \equiv
 \frac{(\zeta\1-\tilde\M_1)(\zeta\1-\tilde\M_2)\dots(\zeta\1-\tilde\M_{2n})}
      {\pi_{2n}(\zeta)},
\end{equation}
where we have defined the polynomial $\pi_{2n}=(\zeta-K_1)\dots(\zeta-K_{2n})$ of degree $2n$ and the matrices
\begin{equation}\label{eq:defMtilde}
 \tilde\M_j:=K_j\1-\frac{(-1)^j}{2\ii\omega_j}\M_j,\quad j=1,\dots,2n.
\end{equation}
Hence all components of $\N$ are rational functions in $\zeta$. In particular, according to \eqref{eq:Nformula}, the numerators of the diagonal elements are polynomials of degree $2n$, whereas the off-diagonal elements have numerators of lower degrees.

Comparing \eqref{eq:defN} and \eqref{eq:Nformula}, we can now determine the polynomial structure (as functions of $\zeta$) of some combinations of the axis potentials. The results are
\begin{equation}\fl\label{eq:Ernstparts}
 f = \frac{\pi_{2n}}{p_{2n}},\quad
 |\E|^2 = \frac{q_{2n}}{p_{2n}},\quad
 \Phi = \frac{p_{2n-1}}{p_{2n}},\quad
 \Phi\bar\E = \frac{q_{2n-1}}{p_{2n}},\quad
 b = \frac{p_{2n-2}}{p_{2n}},\quad
 |\Phi|^2 = \frac{q_{2n-2}}{p_{2n}}
\end{equation}
in terms of polynomials of the indicated degrees. In particular, $\pi_{2n}$ was explicitly defined above, $p_{2n}$ and $q_{2n}$ are real monic polynomials, $p_{2n-1}$ and $q_{2n-1}$ are complex polynomials, and $p_{2n-2}$ and $q_{2n-2}$ are real polynomials.
Note that we have used the parameter condition \eqref{eq:parcond} to conclude that the coefficients of $\zeta^{2n-1}$ in $p_{2n-2}$ and $q_{2n-2}$ vanish, i.e.\ the degrees of both polynomials reduce to $2n-2$ as indicated.

From \eqref{eq:Ernstparts} we can construct the first Ernst potential, 
\begin{equation}\label{eq:ErnstE}
 \E=f-|\Phi|^2+\ii b=\frac{\pi_{2n}-q_{2n-2}+\ii p_{2n-2}}{p_{2n}}
   =:\frac{Q_{2n}}{p_{2n}}.  
\end{equation}
Hence $|\E|^2=Q_{2n}\overline{Q}_{2n}/p_{2n}^2$.
On the other hand, we also have from \eqref{eq:Ernstparts} that $|\E|^2=q_{2n}/p_{2n}$, which leads to the condition
\begin{equation}
 Q_{2n}\overline{Q}_{2n}=p_{2n}q_{2n}.
\end{equation}
This allows us to conclude that some linear factors in \eqref{eq:ErnstE} cancel and $\E$ simplifies to a rational function of lower degree, which we see as follows.
Since $Q_{2n}$, $\overline{Q}_{2n}$, $p_{2n}$ and $q_{2n}$ are monic polynomials, they can all be decomposed into linear factors of the form $\zeta-c$. Let $k\ge0$ be the number of common linear factors in $Q_{2n}$ and $q_{2n}$, i.e.\ factors that would cancel in \eqref{eq:ErnstE}. Then the remaining $2n-k$ factors of $p_{2n}$ all need to be factors of $\overline{Q}_{2n}$. In addition, $\overline{Q}_{2n}$ could have further $l\ge0$ common linear factors with $p_{2n}$, which also are factors of $q_{2n}$ (in case that  $p_{2n}$ and $q_{2n}$ have common factors). Hence $\overline{Q}_{2n}$ and $p_{2n}$ have $2n-k+l$ common factors. Consequently, $\bar\E=\overline{Q}_{2n}/{p_{2n}}$ would effectively be a rational function with numerator and denominator of degree $2n-(2n-k+l)=k-l$. Of course, this must equal the degree of $\E$, which is $2n-k$. From $k-l=2n-k$ we obtain
$k=n+l/2\ge n$. Hence (at least) $n$ linear factors in the formula for $\E$ cancel, and we effectively obtain on $\A_1$
\begin{equation}\label{eq:E}
 \E=\frac{\pi_n}{r_n},
\end{equation}
where $\pi_n$ and $r_n$ are complex, monic polynomials of degree $n$. (These could still have some common linear factors, which would lead to a further reduction of the degree; but this would likely correspond to a spacetime with less than $n$ black holes and hence not be relevant.)

Next we analyse the structure of the second Ernst potential $\Phi$. According to the equation $\Phi=p_{2n-1}/p_{2n}$ [cf.~\eqref{eq:Ernstparts}], we have $|\Phi|^2=p_{2n-1}\bar p_{2n-1}/p_{2n}^2$. Comparing with $|\Phi|^2=q_{2n-2}/p_{2n}$ [also from  \eqref{eq:Ernstparts}], we arrive at the condition
\begin{equation}
 p_{2n-1}\bar p_{2n-1}=q_{2n-2}p_{2n}.
\end{equation}
From a similar discussion of linear factors as above, it follows that $p_{2n-1}$ and $p_{2n}$ have (at least) $n$ common factors. Consequently,  the formula for $\Phi$ on $\A_1$ simplifies to
\begin{equation}\label{eq:Phi}
 \Phi=\frac{\pi_{n-1}}{R_n}
\end{equation}
with complex polynomials of degrees $n-1$ and $n$, respectively, where $R_n$ is a monic polynomial.

Finally, we show that the polynomials $r_n$ and $R_n$ are identical. From \eqref{eq:E} and \eqref{eq:Phi} we obtain $\Phi\bar\E=\pi_{n-1}\bar\pi_n/(R_n\bar r_n)$, but we also have from \eqref{eq:Ernstparts} that $\Phi\bar\E=q_{2n-1}/q_{2n}$, whence $p_{2n}=R_n\bar r_n$. Moreover, $|\Phi|^2$ equals $\pi_{n-1}\bar\pi_{n-1}/(R_n\bar R_n)$ according to \eqref{eq:Phi}, and $q_{2n-2}/p_{2n}$ according to \eqref{eq:Ernstparts}. This implies $p_{2n}=R_n\bar R_n$. Comparing the two formulae for $p_{2n}$, we see that $r_n=R_n$ indeed holds.

Therefore, the final result is the following rational structure for the Ernst potentials on $\A_1$, 
\begin{equation}\label{eq:Ernstformulae}
 \E(\zeta)   = \frac{\pi_n(\zeta)}{r_n(\zeta)},\quad
 \Phi(\zeta) = \frac{\pi_{n-1}(\zeta)}{r_n(\zeta)}
\end{equation}
in terms of complex polynomials, where $\pi_n$ and $r_n$ are monic polynomials of degree $n$, and $\pi_{n-1}$ is a polynomial of degree $n-1$ (or lower --- for example, vacuum configurations with uncharged black holes correspond to $\Phi\equiv 0$ and hence $\pi_{n-1}\equiv 0$).

As special cases, these equations contain the correct form of the Kerr-Newman axis potentials (for $n=1$) and the most general form for possible two-black-hole configurations as derived in \cite{Hennig2019}.

\section{Discussion\label{sec:discussion}}

The discussion of the linear problem, which is equivalent to the Einstein-Maxwell equations in form of the Ernst equations via its integrability condition, has provided us with the most general form of the Ernst potentials for $n$-black-hole equilibrium configurations on the upper part of the symmetry axis. If any regular such configurations exist, then the corresponding axis values of the Ernst potentials necessarily have the rational form \eqref{eq:Ernstformulae}. This means that future investigations of existence of physically reasonable $n$-black hole configurations can concentrate on studying properties of a well-defined finite family of candidate solutions, characterised by those axis data. In principle, it is straightforward, for any given $n$, to obtain the corresponding (unique) solution everywhere off the axis by applying techniques like ``Sibgatullin's integral method'' \cite{Sibgatullin1984,MankoSibgatullin1993}. Generally, this requires  solving linear integral equations with kernels determined by the axis data, but in case of rational boundary values, the problem effectively reduces to solving algebraic equations. In practice, however, these calculations can still be lengthy and involved. Nevertheless, future discussions of these solutions will hopefully reveal whether there are any subsets of the parameter spaces for which the configurations are regular everywhere, i.e.\ free of any pathologies like naked singularities, magnetic monopoles and struts.

 


\end{document}